**Designing of a magnetodielectric system in hybrid organic-inorganic framework, a perovskite layered phosphonate MnO$_3$PC$_6$H$_4$-*m*-Br.H$_2$O**

*Tathamay Basu\*, Clarisse Bloyet, Felicien Beaubras, Vincent Caignaert, Olivier Perez, Jean-Michel Rueff, Alain Pautrat and Bernard Raveau\**

Normandie Univ, ENSICAEN, UNICAEN, CNRS, CRISMAT, 6 Bd du Maréchal Juin, 14050 Caen Cedex, France.

*E-mail: tathamaybasu@gmail.com; bernard.raveau@ensicaen.fr

*Jean-François Lohier*

Normandie Univ, ENSICAEN, UNICAEN, CNRS, LCMT, 6 Bd du Maréchal Juin, 14050 Caen Cedex, France.

*Paul-Alain Jaffrès, Hélène Couthon*

CEMCA UMR CNRS 6521, Université de Brest, IBSAM, 6 Avenue Victor Le Gorgeu, 29238 Brest, France.

*Guillaume Rogez, Grégory Taupier and Honorat Dorkenoo*

IPCMS, UMR Unistra-CNRS 7504, 23 rue du Loess, BP 43, 67034, Strasbourg Cedex 2, France.

**Abstract:**

**Very few hybrid organic-inorganic frameworks (HOIF) exhibit direct coupling between spins and dipoles and also restricted to particular COOH-based system. We show how one can design a hybrid system to obtain such coupling based on the rational design of the organic ligands. The layered phosphonate, $MnO_3PC_6H_5 \cdot H_2O$, consisting of perovskite layers stacked with organic phenyl layers, is used as a starting potential candidate. To introduce dipole moment, a closely related metal-phosphonate, $MnO_3PC_6H_4$-$m$-$Br \cdot H_2O$ is designed. For this purpose, this phosphonate is prepared from 3-bromophenylphosphonic acid that features one electronegative bromine atom directly attached on the aromatic ring in meta position, lowering the symmetry of precursor itself. Thus, bromobenzene moieties in $MnO_3PC_6H_4$-$m$-$Br \cdot H_2O$ induce a finite dipole moment. This new designed compound exhibits complex magnetism, as observed in layered alkyl chains $MnO_3PC_nH_{2n+1} \cdot H_2O$ materials, namely, 2D magnetic ordering ~20 K followed by weak ferromagnetic ordering below 12 K($T_1$) with a magnetic field ($H$)-induced transition ~25 kOe below $T_1$. All these magnetic features are exactly captured in $T$ and $H$-dependent dielectric constant, $\varepsilon'(T)$ and $\varepsilon'(H)$. This demonstrates direct magnetodielectric coupling in this designed hybrid and yields a new path to tune multiferroic ordering and magnetodielectric coupling.**

1. INTRODUCTION:

Multiferroic (MF) systems based on the combination of magnetic and electric orders within a single phase have attracted considerable interest in last years, opening large perspectives for applications in the field of spintronics.[1–5] Particularly, the MF materials, which exhibit coupling between magnetic and electric order parameters, are most promising, since they offer a potential for the control of magnetization by an electric field and *vice-versa*.[6,7] Such a material is also in the current focus from fundamental science aspect which can originate many fascinating magneto(di)electric phenomena.[1,3] However, such single materials are rare, as magnetism is related to partially filled *d* or *f* orbitals and ferroelectricity is favorable in compounds with empty *d*-shells ($d^0$).[3] Among various approaches, one promising route seems to be the search for multiferroicity and magnetoelectric coupling in a hybrid organic-inorganic framework (HOIF) due to the modularity of its structure and geometry.[8,9]

The potential of HOIF as multiferroics was initiated for the metal organic frameworks (MOFs) compounds [$Mn_3(HCOO)_6$](solvent) with solvent =$CH_3OH, H_2O, C_2H_5OH$,[10,11], and [$(CH_3)_2NH_2$]M(HCOO)$_3$ with M= Mn, Co, Ni, Fe.[12–15] These compounds exhibit magnetic ordering between 8 K and 35.6 K and show ferro or antiferroelectric order up to 160-170 K.[12,16] A small magnetodielectric (MD) coupling *via* magnetoelastic coupling is reported for [$(CH_3)_2NH_2$]Fe(HCOO)$_3$.[14] However, in all of these MOFs, magnetic and dielectric (ferroelectric) ordering occurs at completely different temperatures, therefore, none of these compounds have been experimentally proved to exhibit a direct electric control of the magnetization in spite of promising theoretical predictions reported for [$(CH_3)_2NH_2$]Cr(HCOO)$_3$ MOF.[17,18] Recently, the canted antiferromagnetic MOF [$CH_3NH_3$]Co(HCOO)$_3$, with a perovskite- like structure, has been shown to exhibit magneto-electric coupling below magnetic ordering ($T_N$=7.2 K), though no clear dielectric/ferroelectric feature at the onset of magnetic ordering was evidenced.[19] Thus, one of the biggest drawbacks in so-called reported MOFs is that there is no direct one-to-one correlation between temperature and field dependent magnetism and dielectric properties.

The main advantage of hybrid organic-inorganic frameworks is the possibility to control, *via* a careful choice of the organic precursor, the flexibility, topology, and possibly the polarity of the final material. Moreover, the possibility to use a variety of different types of functional ligands and metal ions to tune different functional properties, such as magnetic and electric order, open many possibilities.[20] However, those available reports on multiferroicity in metal-organic framework are restricted to particular systems, that is, 3D perovskite structure containing bridging formate (HCOO) groups in functional ligands. Only very recently, possible multiferroicity and MD coupling is reported in a different metal-organic network apart from so-called COOH-based MOF systems, that is, in a tetragonal framework of $[Co(C_{16}H_{15}N_5O_2)]$.[21] Although, magnetic and electric orderings occurs at a different temperature for this system and no direct one-to-one correspondence is observed between isothermal $M(H)$ and $\varepsilon'(H)$;[21] this result opens up the possibility to design a suitable hybrid framework to test the direct spin-dipole coupling (with one-to-one correlation) apart from reported COOH-based system so far.

A second class of attractive multiferroic hybrids concerns the layered systems, which are of interest, since they allow to combine the flexibility of the framework with robust magnetic and electric properties. In other words, in such structures, the magnetic ordering originates from the inorganic layers containing the transition metal, while the ligands of the organic layers are responsible for electric order (e.g. ferroelectricity). This is the case of bimetallic oxalates[22] as exemplified by the Cr-Mn compound which is ferromagnetic below 3.9 K and might be ferroelectric but no magnetodielectric coupling is documented. The layered perovskite chlorides are also an attractive family as exemplified by the $CuCl_4$-hybrids $CuCl_4(C_6H_5CH_2CH_2NH_3)$ and $CuCl_4(CH_2CH_2NH_3)$, which show a coexistence of ferromagnetic ordering (Tc~9 to 13 K) originating from inorganic $CuCl_4$ layers and ferroelectric order (Tc~247 to 340 K) originating from the H-bonds of the organic layers.[23,24] Although electrical polarization and magnetism are both linked to the buckling of $CuCl_6$ octahedra, no MD effect has been observed to-date in these systems. Recently, layered cobalt hydroxide phosphonates have been reported to exhibit a MD effect. Yet the fact is that this system exhibits MD coupling from short-range magnetic correlation in paramagnetic region.[25]

Thus the investigations of organic-inorganic layered hybrids show that these materials appear promising, though, it is yet to be addressed how one can design a hybrid network where direct MD coupling, *i.e.* a one-to-one correspondence between spins and dipoles, can be achieved by tuning functional ligands, and then one can further think to tune its properties. On such magneto(di)electric material, dielectric constant/ polarization should trace all the magnetic features, that is temperature dependent magnetic features in the absence and presence of magnetic field and magnetic field dependent features at constant temperatures.

Here, we have considered the potential of the structural family of layered (2D) phosphonates, namely the phenyl phosphonate $MnO_3PC_6H_5.H_2O$ which non-centrosymmetric P$mn2_1$ structure consists of single oxygenated perovskite layers interconnected through organic phenyl layers[26]. Such a compound which was previously shown to exhibit antiferromagnetic (AFM) properties ($T_0$=20 K) [27] can then be considered as initial potential candidate for the realization of possible multiferroic /magnetoelectric properties. For this purpose, we aimed to introduce on the aromatic rings an electron-withdrawing atom in order to induce a polarization of the organic moiety. Bearing in mind that the polarizability of the atom could be an important feature for the final properties of the dipoles we selected to add a bromine atom which is a compromise between electronegativity and polarizability. In this way, we specifically introduced a dipole moment in the organic layers by substituting one hydrogen atom by one bromine atom on the phenyl ring, leading to the closely related layered phosphonate $MnO_3PC_6H_4$-*m*-$Br.H_2O$. Of note, we placed the bromine atom in *meta* position (relative to the phosphorus atom) in order to limit the symmetry of the organic precursor.

In this study, we describe the 2D antiferromagnetic (AFM) ordering ($T_0$=20 K) for the new designed compound similar to the previous one (without Br)[27] and we show that this new phosphonate behaves as a weak ferromagnet or canted antiferromagnet (CAFM) at lower temperature ($T_1$=12 K), similar to the layered alkyl phosphonates $MnC_nH_{2n+1}PO_3.H_2O$.[28] Importantly, we demonstrate the possibility of designing magnetodielectric coupling for these hybrid materials. Also, we document the fascinating phenomenon of magnetic field-induced "meta-magneto-electric-type" transition, which is rarely reported.

## 2. RESULTS AND DISCUSSION:

### 2.1. Synthesis of the 3- bromophenylphosphonic acid, organic precursor of the hybrid Mn(H$_2$O)PO$_3$-C$_6$H$_4$-*m*-Br:

The synthesis of the layered phosphonate Mn(H$_2$O)PO$_3$-C$_6$H$_4$-*m*-Br requires in a first step the preparation of the 3-bromophenylphosphonic acid. Aromatic phosphonic acids can be prepared by a series of methods[29] but one of the most convenient consists in a phosphonation of bromoaryl derivatives[30] according to the Tavs's method[31] or a modified version of this catalytic assisted Arbuzov reaction.[32] This procedure is usually very efficient for the preparation of mono[33] or poly aryl-phosphonates.[34] Then, the hydrolysis of phosphonate to phosphonic acid can be achieved in acidic condition.[34] Note that we placed the bromine atom in meta position respective to the phosphonic acid on the benzene ring because we have previously observed that the use of rigid organic precursors that feature a low symmetry contributed to the formation of non-centrosymmetric hybrid materials.[25,35,36]. To this goal, we used an excess of 1,3 dibromobenzene as substrate to favor the monophosphonation to produce compound **1** (Scheme 1). Then the hydrolysis of the phosphonate with HCl produced 3-bromophenylphosphonic acid **2** in high yield. In a second step, the reaction of the latter with manganese nitrate hydrate in presence of urea allowed the layered phosphonate Mn(H$_2$O)PO$_3$-C$_6$H$_4$-*m*-Br to be obtained as a pure phase.

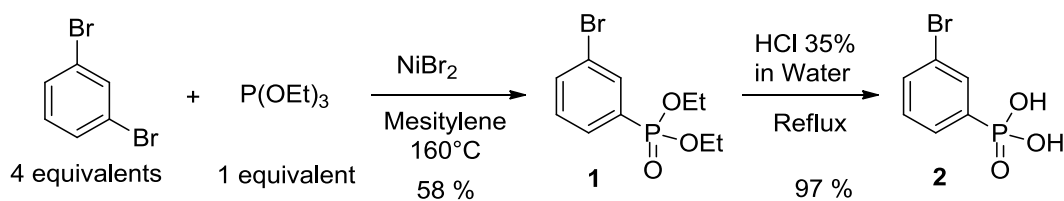

**Scheme/ Equation 1:** Synthesis of phosphonic acid **2** from 1,3-dibromobenzene.

### 2.2. Structural characterization

The powder X-ray diffraction (PXRD) pattern of the phosphonate MnO$_3$PC$_6$H$_4$-*m*-Br.H$_2$O (**Figure 1**) shows the good crystallization of the sample and its high purity. It can be indexed in the orthorhombic non centrosymmetric P*mn2$_1$* space group with the same parameters as MnO$_3$PC$_6$H$_5$.H$_2$O, i.e. a~5.73Å b~14.33 Å c~4.94 Å, indicating that both structures are closely related.[26] However, X-ray data collected from single crystals of MnO$_3$PC$_6$H$_4$-*m*-Br.H$_2$O (**Table 1**) evidenced a doubling of the periodicity along **b**. Non-linear optical measurements indicated that this phase is non centrosymmetric [see inset of Figure 1]. The poor quality of the crystals of the Br-substituted phase hindered an accurate structure determination. Although no deviation from 90° was evidenced for the α, β and γ angles, the structure could only be solved in monoclinic symmetry using the space group P1c1 ; the reliability factor R did not fall below 0.125 due to the poor quality of the crystals showing stacking faults and twinning. Nevertheless, a structural model could be established where the atomic positions of Mn, O, P and C forming infinite [MnPO$_3$(H$_2$O)C] layers similar to those reported by Cao et al.[26] for MnO$_3$PC$_6$H$_5$.H$_2$O, were clearly identified (see **Supplementary Information** Tables S1). Note that a complex disordering of the phenyl rings in the latter compound was observed by these authors.[26] A detailed determination of the structure of these compounds would require further investigations for the growth of larger single crystals with a better quality and is out of the scope of the present study.

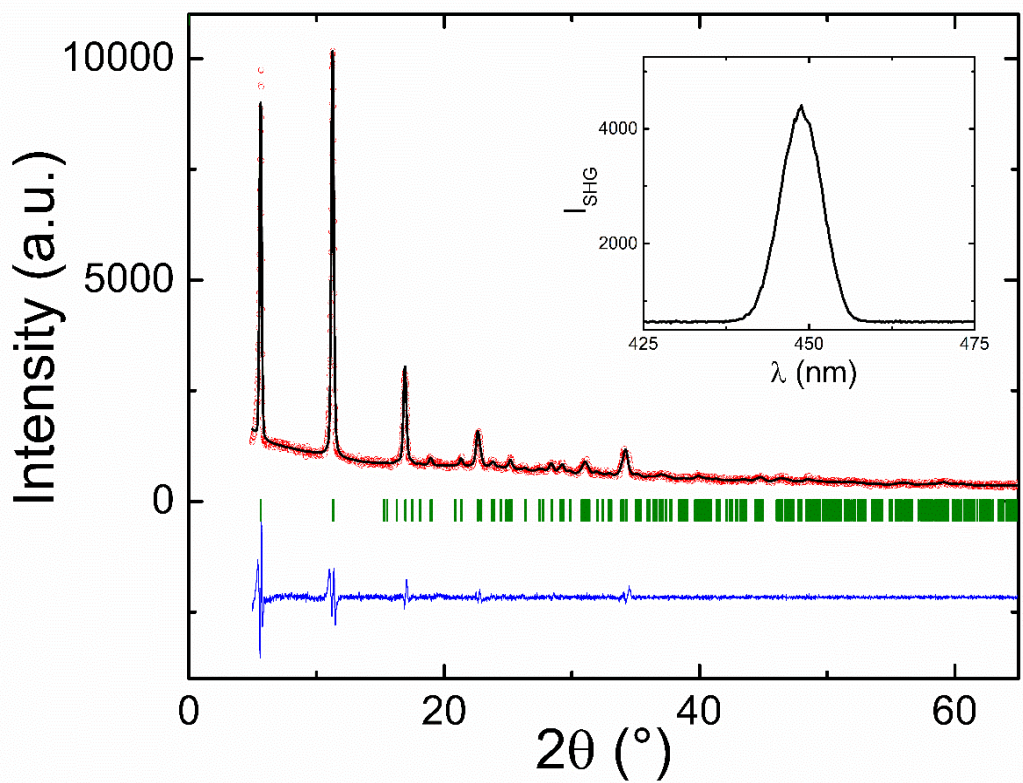

**Figure 1:** X-ray diffraction pattern of MnPO$_3$C$_6$H$_4$-*m*-Br.H$_2$O. Inset shows the non-linear optical (Second Harmonic Generation) results to confirm non-centro symmetry of this compound.

Thus, the structure of the Br substituted phosphonate consists, like MnO$_3$PC$_6$H$_5$.H$_2$O of inorganic (001) MnO$_3$.H$_2$O layers with the perovskite geometry (**Figure 2**a), built up of distorted corner-shared MnO$_5$(H$_2$O) octahedra. However the orientation of the layers with respect to each other is different in the two structures: two successive layers correspond through a translation in MnO$_3$PC$_6$H$_5$.H$_2$O (Figure 2b), whereas they are deduced by a *c* mirror plane in MnO$_3$PC$_6$H$_4$-*m*-Br.H$_2$O (Figure 2c). The tetrahedral PO$_3$C phosphonate groups are grafted on these layers, in such a way that each PO$_3$C group shares its three oxygen atoms with four Mn octahedra, contributing to the rigidity of the MnO$_3$.H$_2$O layers. The phenyl groups lie above and below these layers with various possible orientations.

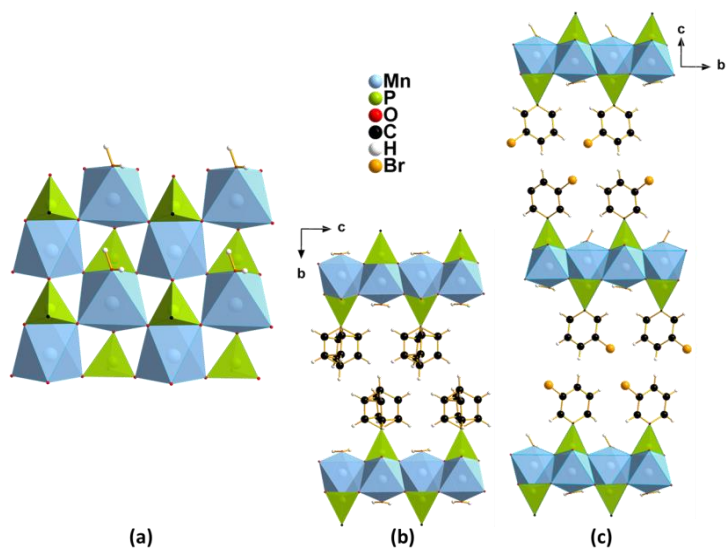

**Figure 2: Structures of MnPO$_3$C$_6$H$_5$.H$_2$O and MnPO$_3$C$_6$H$_4$-*m*-Br.H$_2$O: (a) perovskite-like MnPO$_3$H$_2$O layer containing PO$_3$C tetrahedra, (b) view of the MnPO$_3$C$_6$H$_5$.H$_2$O structure along $\vec{a}$ showing the $b \approx 14$ Å periodicity after Ref. 25 (c) view of the MnPO$_3$C$_6$H$_4$-*m*-Br.H$_2$O structure along $\vec{a}$ showing the c ≈ 31 Å periodicity.**

**Table 1:** Crystallographic data of Mn(H$_2$O)PO$_3$-C$_6$H$_4$-*m*-Br obtained by single crystal X-ray diffraction at 150 K

| Formula | Mn(H$_2$O)PO$_3$-C$_6$H$_4$-*m*-Br |
|---|---|
| FW (g/mol) | 307.94 |
| Space group | P1c1 |
| a (Å) | 4.8875(5) |
| b (Å) | 5.7882(7) |
| c (Å) | 31.043(4) |
| α (°) | 90 |
| β (°) | 90 |
| γ (°) | 90 |
| Z | 4 |
| V (Å$^3$) | 878.21(17) |
| d$_{calc}$ (g/cm$^3$) | 2.32885 |
| μ (mm$^{-1}$) | 6.209 |
| radiation source λ (Å) | Mo K$_α$ = 0.71073 |
| Pattern range 2Θ (°) with F$^2$ > 2σ(F$^2$) | 5.248 – 54.362 |
| no. of reflexions | 2675 |
| no. of soft constraints | 2 |
| Weighted R factor | 0.3045 |
| R[F$^2$ > 2σ(F$^2$)] | 0.1289 |
| R$_{int}$ (internal R-value) | 0. 0464 |
| S (Goodness of the fit) | 1.230 |

## 2.3. Magnetic behavior of MnO$_3$PC$_6$H$_4$-*m*-Br.H$_2$O

The dc susceptibility (here, $\chi$= *M/H*) of the phosphonate MnO$_3$PC$_6$H$_4$-*m*-Br.H$_2$O as a function of temperature is shown in figure 3a for a magnetic field of 1 kOe in zero-field-cooled and field-cooled (FC) conditions. The $\chi(T)$ exhibits an antiferromagnetic (AFM) type broad peak around 20 K ($T_0$), as previously observed for the compound $MnPO_3C_6H_5 \cdot H_2O$ by Le Bideau et al. [27]. The Curie-Weiss fit of the inverse susceptibility for $T > 150$ K yields an effective moment of $\mu_{eff} = 5.77$ $\mu_B$ and a Curie-Weiss temperature of $\Theta_{CW} = -35.4$ K, with a temperature independent contribution of $\chi_0 = -1.4*10^{-4}$ emu/mol. The large diamagnetic contribution from organic ligand yields such a large value of $\chi_0$, which is typical for an organic-based system.[37] The paramagnetic Curie-Weiss behavior deviates from linearity below ~70 K which suggests existence of short-range magnetic correlation far above $T_0$. The effective moment nearly agrees with the theoretically obtained value for spin only moment of $Mn^{+2}$ (5.92 $\mu_B$ for $S=5/2$). The negative sign of $\Theta_{CW}$ denotes the presence of AFM exchange interaction. However, $|\Theta_{CW}| > T_0$, and further, a closer look reveals that the peak around 25 K is quite broad; these two features are common characteristic of a magnetically frustrated system, either geometrically frustrated (like, triangular arrangement of spin) or dimensionally frustrated (say, 2D frustration) or exchange frustrated (due to different ferromagnetic and antiferromagnetic competing interaction). Here, the degree frustration ($|\Theta_{CW}/T| \sim 2$) is nominal. Bearing in mind that the structure of this phosphonate consists of single perovskite layers separated by double organic layers that are ~15 Å thick, it is quite clear that the inter-planar distance along the "a" direction is much longer than Mn-Mn distance in "b-c" plane. Therefore, this feature can be described as 2D magnetism resulting from dominating antiferromagnetic in-plane exchange interaction and negligible inter-planar exchange interaction.

Importantly, another feature appears below ~ 12 K ($T_1$) on the $\chi(T)$ curve (figure 3a) which was missed by Le Bideau et al,[27] for $MnPO_3C_6H_5 \cdot H_2O$ but was reported by Carling et al.[28] in the case of the alkyl phosphonates, $MnO_3PC_nH_{2n+1} \cdot H_2O$, a similar layered structure. The possible canted antiferromagnetic ordering was speculated through a preliminary magnetic investigation.[28] Here we will discuss the magnetization result in detail as a function of temperature and magnetic field. The latter (feature below 12 K) is much better evidenced for a magnetic field of 100 Oe, in which $\chi(T)$ exhibits a very sharp peak at ~ 12 K (see inset of figure 3a). Interestingly, the FC curve shoots up at 12 K and exhibits a broad peak around 7 K followed by a slow decrease down to 2K (inset of figure 3a). The $\chi(T)$ curve for 1 kOe also exhibits a similar AFM-type peak below $T_1$ followed by continuous decrease with lowering $T$ in both ZFC-FC conditions, though a clear ZFC-FC bifurcation is observed (figure 3a). Such a bifurcation in ZFC-FC is not a typical characteristic of a pure AFM system. This bifurcation of ZFC/FC measurements may originate from magnetic domain anisotropy in ferro (ferri) magnetic system or as a result of competing FM-AFM interaction. The $T$-dependent magnetization in presence of different magnetic fields is plotted in figure 3b. For high magnetic field, H ≥ 10 kOe, no ZFC-FC bifurcation is observed. Interestingly, no sharp feature is observed below 12 K for 10 kOe magnetic field, though $\chi(T)$ exhibits a weak kink at 12 K and then continuously decreases with lowering $T$ down to 2 K. For higher magnetic field ($H=$ 30kOe), the $\chi(T)$ exhibits a weak peak below ~12 K and then continuously increases below 10 K with lowering the temperature down to 2 K (see figure 3b), unlike an AFM system. By increasing the magnetic field further up to 50 kOe, the magnetization continuously increases below $T_1$. The continuous increase of magnetization with lowering temperature below magnetic ordering is one of the typical characteristics of a ferromagnetic-type system. Nevertheless, the increase is not as sharp as observed in a pure FM system, but rather shows a slow continuous evolution as observed in a weak ferromagnet or canted AFM -type system. In addition, the magnetic ordering temperature $T_1$ slightly shifts towards higher value by increasing the magnetic field, i.e., from 12 K for 10 kOe to 13 K for 50 kOe (see figure 3b), which is consistent with ferromagnetism.

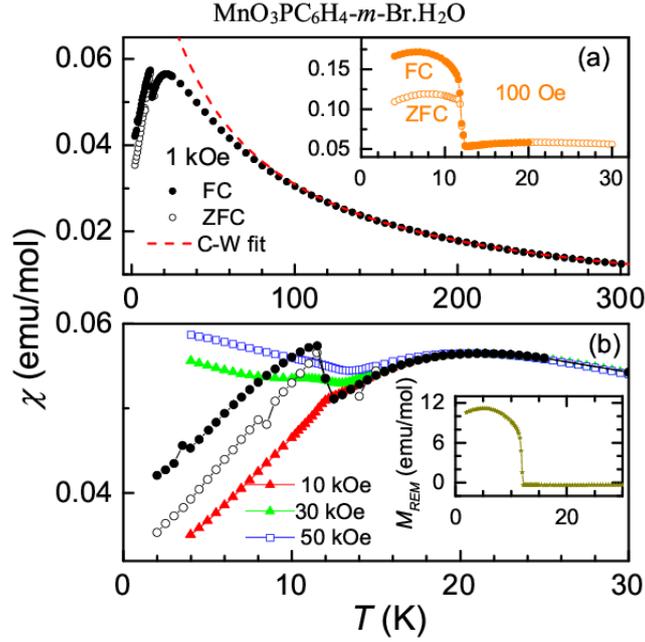

**Figure 3:** (a) Dc susceptibility as a function of temperature for 1 kOe magnetic field for zero-field-cooled (open symbol) and field-cooled (closed symbol) condition from 2-300 K for the compound $MnO_3PC_6H_4$-*m*-$Br.H_2O$; Curie-Weiss fitting in paramagnetic region (150-300 K). Inset of (a): dc susceptibility as a function of *T* for 100 Oe for ZFC (open symbol) and FC (closed symbol) conditions from 2-30 K. (b) *T*-dependent dc susceptibility in presence of different magnetic field (10, 30 and 50 kOe) from 2-30 K, ZFC-FC curves superimposes. Inset of (b): Remnant magnetization is plotted as a function of *T*, as described in the text.

To confirm the nature of magnetic ordering, we have also performed remnant magnetization ($M_{REM}$) as a function of temperature (shown in inset of figure 3b). This can give better information about the nature of magnetism, which is not documented earlier. The $M_{REM}$ measurement has been performed in the following condition: First, the sample was zero magnetic field cooled down to the lowest temperature of 2 K, second, a magnetic field of 10 kOe was applied for 1 min at 2 K and then was decreased to zero, finally dc magnetization was measured vs *T* at increasing temperature. The $M_{REM}(T)$ shows a broad feature (peak) ~ 7 K, followed by a sharp drop at 12 K with increasing temperature and the magnetization value becomes zero above 12 K within the resolution of the instrument. This behavior supports the appearance of ferro (ferri) magnetic-type ordering at $T_1$.

Isothermal magnetization $M(H)$ data for selected temperatures are shown in figure 4 up to 50 kOe. A very weak (negligible) magnetic hysteresis is observed at low $T$ (see left inset of figure 4a and 4b for $T=2$ and 7 K, respectively), which is consistent with the presence of ferromagnetic interaction for $T<T_1$. The $M(H)$ linearly increases with increasing $H$ up to 20 kOe, then exhibits a weak jump around 25 kOe and then further increases linearly above 40 kOe for $T<T_1$ (see figure 4a and 4b for $T=2$ and 7 K respectively). The jump ~ 25 kOe is clearly depicted as a sharp peak in first order derivative of $M(H)$, as shown in the right inset of figure 4a and 4b for $T=2$ and 7 K respectively. Such a jump in $M(H)$ is normally observed in case of $H$-induced meta-magnetic transition. However, we did not observe any hysteresis at the onset of meta-magnetic transition as a characteristic of first-order transition; this suggests that we missed the hysteresis due to the very weak nature of the meta-magnetic transition. The magnetization at low $T$ does not saturate even at a very high magnetic field (say, 50 kOe). Also, the value of $M$ at very high $H$ (say, M @50 KOe, 2K) ~ 0.5 $\mu_B$/Mn is much less (10 orders less) than the Mn$^{+2}$ saturation moment (M$_S$ ~ 5 $\mu_B$) for spin-only value of $S=5/2$ and Lande-$g$-factor for spin-value $g_s=2$. Eventually, the $M(H)$ curve varies linearly with $H$ (even at very high $H$). This linear increase in $M(H)$ is not ideal for a ferromagnet, rather, it indicates towards AFM interaction. Thus, our results confirm presence of both FM and AFM interactions in this system below $T_1$. The ZFC-FC bifurcation in $\chi(T)$ is also aligned with this conclusion. The isothermal $M(H)$ varies linearly for $T>T_1$, (see figure 4c for 15 and 35 K). No hysteresis and no meta-magnetic-type feature (jump around 25 kOe) is observed in isothermal $M(H)$ for $T>T_1$ (see figure 4c and inset for 15 K). This is consistent with 2D AFM and paramagnetic behavior at such high temperatures.

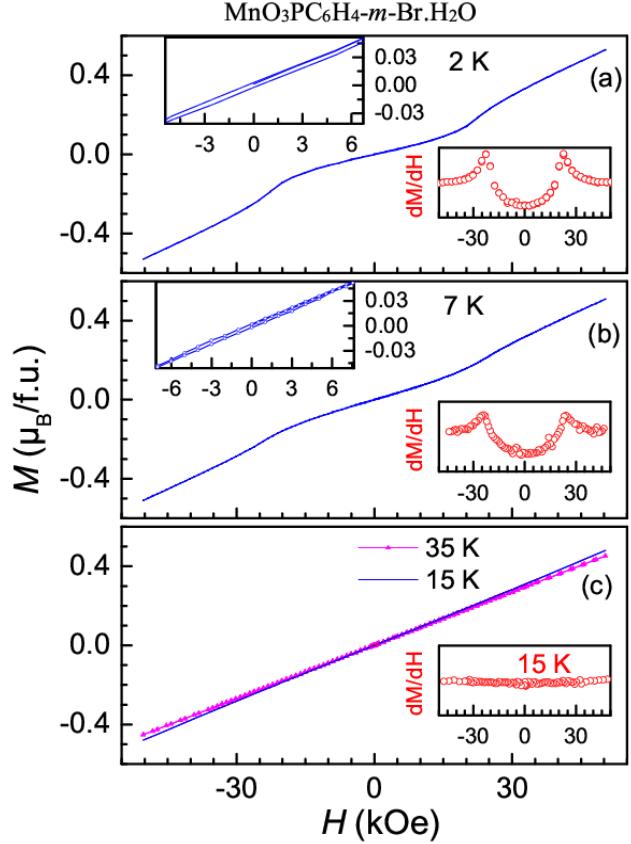

**Figure 4:** Isothermal magnetization as function of magnetic field for $T=$ 2 K (a), 7 K (b), 15 and 35 K (c), for the compound $MnPO_3C_6H_4$-$m$-$Br.H_2O$. Right inset of (a), (b) and (c) shows $dM/dH$ as a function of $H$ for 2, 7 and 15 K, respectively, to highlight the metamagnetic transitions around 25 kOe for $T<T_1$. Left inset of (a) and (b) shows the enlarged version of $M(H)$ at low magnetic field to highlight the weak hysteresis at 4 and 7 K, respectively.

The existence of both FM and AFM interactions in this phosphonate allows a possible picture of its magnetic structure to be proposed. The intra-plane interactions are antiferromagnetic and yield a 2D-AFM ordering around $T_0$, the broad peak at $T_0$ with negative $\Theta_P$ vouches for that. With further decreasing temperature, a weak long-range magnetic ordering (3D ordering) is established below $T_1$ due to development of ferromagnetic exchange interactions between the planes. Such a scenario of long range ferromagnetism mediated by interlayer organic spacers and resulting from dipolar interactions has been previously shown for layered hybrid organic-inorganic compounds by Drillon and co-workers.[38–40] Therefore, the magnetic ground state of this system could be characterized as a weak ferromagnet as a result of anisotropic FM and AFM interaction arising from intra-plane and inter-plane exchange interactions. Application of high magnetic field tries to reorient all the spins, a field induced transition around 25 kOe is observed. However, complete ferromagnetism is not developed even at very high magnetic field, otherwise, the *M(H)* would tend towards saturation rather than linear increase. The continuous linear increase in *M(H)* up to very high *H* is very-well consistent with the canted AFM where the canting angle decreases with increasing *H*. Probably the application of a magnetic field stabilizes the ground state by minimizing the FM-AFM interaction with an (in-plane) canted AFM structure. Canted AFM structures are also speculated in some MOF systems.[21,41]

The weak ferromagnetism (or, canted AFM) is quite compatible with the existence of Dzyaloshinskii-Moriya (D-M) interactions in a non-centro symmetric system. The presence of competing FM-AFM interactions favors lattice distortion to minimize the magnetic energy *via* a stabilized magnetic structure. We have also specifically designed this system to be electrically polar by introducing bromine (Br) in the phenyl ring, as the electron withdrawing properties of Br and its polarizability induce local dipole moments. Therefore, this kind of complex magnetic system with non-centrosymmetric crystal structure is a potential candidate to investigate magnetodielectric coupling.

**2.4. Dielectric and Magnetodielectric behavior of MnO$_3$PC$_6$H$_4$-*m*-Br.H$_2$O**

The real part of dielectric constant *(ε′(T))* of MnO$_3$PC$_6$H$_4$-*m*-Br.H$_2$O is plotted as a function of temperature from 5-50 K for a fixed frequency of 71.4 kHz in presence of different magnetic fields (0, 10, 50, 100 kOe), as shown in **Figure 5**a. The low value of loss tangent

(tanδ<<1, see inset of Figure 5a) confirms the highly insulating nature of this system, which is a perquisite for dielectric and magnetodielectric measurements. A clear broad peak is observed in $\varepsilon'(T)$ around $T_0$ in presence of zero magnetic field, exactly tracing the magnetic feature. A change of slope in $\varepsilon'(T)$ is observed below 12 K for zero magnetic field, which is attributed to a magnetic feature at $T_1$. Application of a 10 kOe-magnetic field has no significant effect on the behavior of $\varepsilon'(T)$ compared to that for zero magnetic field. The application of a high magnetic field of 50 kOe has negligible effect at $T_0$ (i.e., the peak value in $\varepsilon'(T)$ at $T_0$ slightly increases). However, a weak but clear feature is observed at the onset of $T_1$. The application of further high magnetic field (100 kOe) slightly increases the peak value at $T_0$ and captures the magnetic feature at $T_1$. This result indicates the presence of magnetodielectric coupling and a direct one-to-one correspondence between magnetic and dielectric feature in the magnetic $T$-regime. We have further confirmed this MD coupling through $H$-dependent dielectric constant as well. The fractional change in dielectric constant ($\Delta\varepsilon'/\varepsilon' = [\{\varepsilon'(H)-\varepsilon'(0)\}/\varepsilon'(0)]$) as function of $H$ is documented in Figure 5b for different temperatures. For $T<T_1$ (say, $T=7$ K), the dielectric constant increases with increasing $H$ and exactly traces the $H$-induced meta-magnetic-type transition around 25 kOe and further increases quadratically with $H$. For $T_1<T<T_0$ (say, $T=15$ K), the dielectric constant quadratically increases with increasing $H$, obviously, no feature around 25 kOe is observed. The ME coupling above $T_0$ is zero (see figure 5b for 40 K). The $H$-induced feature around 25 kOe in $\Delta\varepsilon'(H)$ is termed as 'meta-magnetodielectric-type' transition which is attributable to meta-magnetic-type transition in $M(H)$. The meta-magnetic-type feature in $M(H)$ is strong at 2 K but very weak at 7 K, interestingly, our dielectric results clearly capture this transition even at 7 K due to presence of strong ME coupling. This result reveals the merit of the dielectric measurements to capture such a weak $H$-induced magnetic transition, as reported in many multiferroic and magnetoelectric oxide compounds (e.g. Ref.[42] and references there in). This kind of meta-magneto-electric-type behavior is observed in MOF system [CH$_3$NH$_3$]Co(HCOO)$_3$.[19] All the features in $\varepsilon'(T)$ and $\Delta\varepsilon'(H)$ are repeated for other frequencies as well (therefore, not shown here). It is to be noted, that a direct one-to-one correlation between magnetism and dielectric in both $T$-and $H$-dependent features, as observed in the title compound, is rarely reported in HOIF systems.

Here, we will speculate the possible mechanism of ME coupling. A clear peak is observed at the onset of 2D magnetic ordering. Such a peak in dielectric constant around $T_0$, mimicking the low dimensional magnetism, could be attributable to dipolar ordering as a result of minimization of overall ground state energy, which is recently documented in an one-dimensional helical-chain system in metal-organic hybrid network.[43] Low-dimensional magnetism favored ferroelectricity and magnetoelectric coupling from short-range magnetic correlation has been reported in frustrated/ spin-chain system.[44–48] The small influence of $H$ on dielectric constant around $T_0$ could be *via* magneto-elastic coupling. However, a strong influence of $H$ on the dielectric feature below $T_1$ suggests that another mechanism (apart from magneto-elastic coupling) should exist at low temperature. The title system magnetically orders below $T_1$ with a canted AFM structure, which can favor ME coupling via D-M interaction. The competing anisotropic magnetic interactions are released by a stabilized magnetic structure below $T_1$ which distorts the lattice (lattice constant, bond-length, etc.), and in turn, affects the dielectric constant. Obviously, the magnetic field has a strong influence on the magnetic structure and thus, on the dielectric constant due to ME coupling, as observed in the feature of $\varepsilon'(T)$ and in $\varepsilon'(H)$ below $T_1$. Here, in this weak ferromagnet (canted antiferromagnet), ME coupling below $T_1$ may occur through D-M interaction. The role of D-M interaction from canted AFM structure on ME coupling is indeed reported in MOF [CH$_3$NH$_3$]Co(HCOO)$_3$.[19] One of the most well-known multiferroic ME system RMn$_2$O$_5$ (R=rare-earth) exhibits strong ME coupling where both D-M interaction and exchange-striction are documented, on which magneto-elastic coupling also has significant role.[49–51] The MD coupling may also arise from magneto-structural effect or any extrinsic effect like leakage. Nevertheless, the highly insulating nature of this sample excludes any leakage contribution. We have also performed dieletric and magnetodielectric investigation of the parent phosphonate compound $MnO_3PC_6H_5.H_2O$, which does not contain Br-atom. Preliminary investigations confirmed that the parent phosphonate structure $MnO_3PC_6H_5.H_2O$ presents a magnetic behavior absolutely similar to that of the title compound, but no dipolar ordering or ME coupling was detected (See **Supplementary Information**). The absence of MD coupling in the parent phase excludes the possibility of magneto-structural effect. The latter would indeed induce MD coupling in both phosphonate systems.

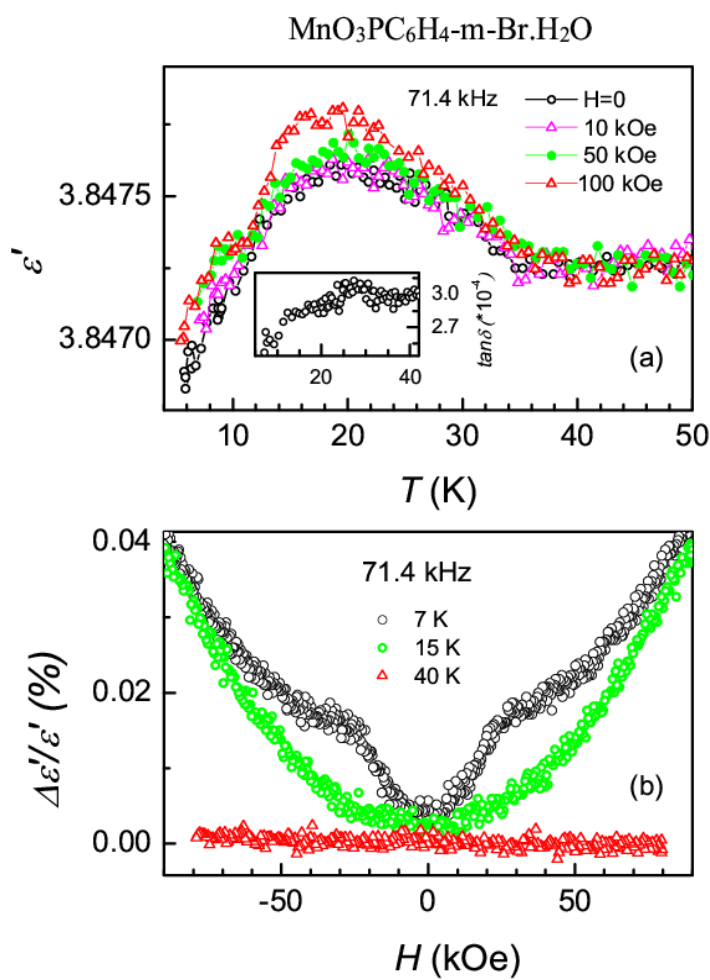

**Figure 5:** (a) Real part of dielectric constant as a function of temperature from 5-50 K for a fixed frequency of 71.4 kHz in presence of different magnetic fields (0-100 kOe) for the compound $MnO_3PC_6H_4$-*m*-$Br.H_2O$. Inset: shows loss tangent for zero magnetic field. (b) Fractional change in real part of dielectric constant as a function of magnetic field at selective temperatures for a fixed frequency of 71.4 kHz.

Now, we will discuss the significance of this kind of layered hybrid metal-organic system and the effect of introduction of an electronegative atom into the parent phosphonate structure $MnO_3PC_6H_5.H_2O$ to tune multiferroic ordering and magnetodielectric coupling. One advantage of this layered structure is that its magnetism is before all controlled by the interlayer distance and spin correlations within the inorganic layers (see ref 39-40 and ref. therein). Therefore, magnetic ordering can be tuned by modifying the organic ligand (i.e. changing the inter-planar distance). Its non centro-symmetric structure and canted antiferromagnetism are favorable features for the appearance of ferroelectricity and ME coupling. The absence of MD coupling in $MnO_3PC_6H_5.H_2O$ suggests either a too small value of the dipole moments or a disordered arrangement of the latter. Thus in this scenario, we have introduced a more electronegative Br atom at meta position of the benzene rings.

The introduction of bromine atom is the key point to maintain a similar non-centro symmetric crystal structure and similar magnetic structure and to contribute to permanent local dipole moments for this layer phosphonate. As a consequence, the dipolar ordering observed for $MnO_3PC_6H_4$-*m*-$Br.H_2O$ originates, at least partly, from the electronegativity of Br (2.96 according to the Pauling electronegativity scale) which is slightly higher than those of carbon and hydrogen atoms (respectively 2.55 and 2.10 according to the Pauling electronegativity scale). This may lead to a polarization of the Br-C bond due to high polarizability of bromine atom. This Br-C polarization could further help to order the dipole moments (if already present) in a preferred direction. Moreover, dipole-dipole interactions may appear within the organic layers due to the possible existence of C-H…Br-C hydrogen bonds between the $C_6H_4Br$ rings

(For example, see Ref [52–55] where the role of Br- species and the formation of Br…H bond is discussed to generate electric polarization) and consequently induces local-(electric) dipolar ordering. These dipoles can be arranged in a particular direction at low temperature to minimize the ground state energy, which produces a permanent finite dipole in macroscopic structure, based on the presence of the electro-withdrawing bromine atoms. Such a hypothesis is corroborated by our dielectric and ME results. The meta-magnetodielectric-type transition around 25 kOe clearly suggests that magnetic field has a pronounced effect on electric dipoles. The way the dipoles in organic layers affect the magnetism and *vice-versa* is still a pending question which would require detailed knowledge of the precise structure of this compound. This suggests our proposal that this kind of layered phosphonate structure is itself prone/favorable for dipolar ordering and ME coupling, where slight modification is needed to order the dipoles to a preferred direction to obtain significant polarization, as documented by our results. In particular, improper ferroelectricity in a magnetodielectric system implies very small displacements of the atoms in the inorganic layers. It is to be noted that such a small structural change (displacement of atoms) is rarely traced in the case of an improper ferroelectrics (which is most of the case in magnetodielectric system), because, it cannot be detected through lab-based X-ray facility. This would require synchrotron X-ray investigation at low temperature below long-range ordering to detect such a tiny change in structure or atomic displacements. Microscopic and theoretical investigations are highly warranted to shed more light on magnetism, dielectric and ME coupling in this new class of hybrid ME systems.

## 3. Conclusions

This detailed study of 2D-layered phosphonate, $MnO_3PC_6H_4$-*m*-$Br.H_2O$, demonstrates that this hybrid exhibits 2D magnetic ordering around 20 K and a complex magnetic ordering below 12 K with presence of competing FM-AFM interactions. The latter state is probably stabilized to a canted AFM structure. In contrast to the starting material MnO$_3$PC$_6$H$_5$.H$_2$O, the bromine substituted phosphonate MnO$_3$PC$_6$H$_4$-*m*-Br.H$_2$O shows ME coupling and a fascinating phenomenon of "meta-magneto-electric-type" transition. For this compound, a direct one-to-one correlation between spins and dipoles is established through documentation of both *T* and *H*- dependent magnetic and dielectric results, which is rare in the field of hybrid-organic-inorganic-framework. Thus, starting from a layered phosphonate, MnO$_3$PC$_6$H$_5$.H$_2$O, where ME coupling is negligible, the introduction of a halide atom possessing noticeable electron-withdrawing properties (Br) contributes to induces permanent dipoles in the organic ligand layers, generating ME coupling. Here, we have demonstrated a possible scenario where one can design in a first step, a suitable magnetic framework, and then in a second step, one can generate/enhance magneto-electric coupling by selectively modifying the chemical nature of the organic ligands. Therefore, this system offers a good starting point to design and control the magneto-electric effect and tune multiferroic ordering by modifying the ligand in HOIF systems in general.

## 4. Experimental Section

### 4.1. Synthesis

*4.1.1. Synthesis of the precursor: diethyl 3-bromophenylphosphonate* **1**

NiBr$_2$ (462 mg, 2.1 mmol) and 1,3-dibromobenzene (30.5 g; 129 mmol, 4 eq) suspended in mesitylene (15 mL) was heated to 160°C under nitrogen. A solution of triethylphosphite (5.6 mL, 32.7 mmol) in mesitylene (40 mL) was added by a slow and careful dropwise addition. Then, the reaction was further stirred at 165°C for eight hours. After cooling, mesitylene was removed by distillation under vacuum. The resulting mixture was dissolved in dichloromethane (250 mL) and water (250 mL) was added. The solution was stirred at room temperature during four hours. The two layers were separated and the organic layer was washed with water (2 x

300 mL). The organic layers were dried over MgSO$_4$ and evaporated under reduced pressure to yield clear oil. The excess of the 1,3-dibromobenzene was partly removed by Kugelrohr distillation under vacuum (3.10-2 mbar). The product was then purified by chromatography on silica with dicholoromethane/ethanol (100/2) as eluting solvent. (Spectroscopic characterizations are available in **Supporting Information**)

*4.1.2. Synthesis of the ligand: 3-bromophenylphosphonic acid* **2**

Diethyl 3-bromophenylphosphonate (5.51 g; 188 mmol) and concentrated HCl (35 % in water, 120 mL) was refluxed for 12h. The excess of HCl and water were removed under vacuum and dried by azeotropic distillation with toluene to produce, after drying, 4.32 g of the expected compound (97%). Spectroscopic characterizations are available in **Supporting Information**.

*4.1.3. Synthesis of Mn(H$_2$O)PO$_3$-C$_6$H$_4$-m-Br by classical hydrothermal method*

In a 50 mL Teflon liner, an equimolar mixture of Mn(NO$_3$)$_2$.4H$_2$O (0.085 g, 0.337 mmol, 1 eq), 3-bromophenylphosphonic acid (0.080 g, 0.337 mmol, 1 eq) and urea (NH$_2$)$_2$CO (0.020 g, 0.337 mmol, 1 eq) were dissolved in 10 mL of distilled water. Then, the liner was transferred into a Berghof DAB-2 pressure digestion vessel and heated according to the following thermal cycle: first heating from room temperature to 140°C in 30 hours, kept at 140°C during 30 hours and cooled to room temperature in 36 hours. The final compound, obtained as transparent platelets, was filtered, washed with water, rinsed with ethanol and dried in air. Elemental Analysis (MnPO4C6H6Br): found (%): C 23.81, H 2.49, calc. (%): C 23.40, H 1.96. Thermogravimetric Analysis (Figure S2 in Supporting Information) has been used to confirm the presence of one water molecule in the formula: found (%): 5.35, calc. (%): 5.80.

*4.1.4. Synthesis of Mn(H$_2$O)PO$_3$-C$_6$H$_4$-m-Br by microwave assisted method*

In a 20 mL glass vial, an equimolar mixture of Mn(NO$_3$)$_2$.4H$_2$O (0.053 g, 0.211 mmol, 1 eq), 3-bromophenylphosphonic acid (0.050 g, 0.211 mmol, 1 eq) and urea (NH$_2$)$_2$CO (0.013 g, 0.211 mmol, 1 eq) were dissolved in 10 mL of distilled water. The vial was inserted into a Biotage

Initiator+ microwave apparatus and heated according to the following thermal cycle: first heating from room temperature to 140°C within 1 min, kept at 140°C for 1 hour and finally cooled to room temperature within 1.5 min with pressurized air. The final compound, obtained as powder made of transparent platelets, was filtered, washed with water, rinsed with ethanol and dried in air.

### 4.2 Single crystal and powder X-ray diffraction and analysis

Crystallographic data sets were collected from single crystal samples. Collections were performed using a Bruker Kappa APEXII CCD diffractometer. The initial unit cell parameters were determined by a least-squares fit of the angular setting of strong reflections, collected by a 6.0° scan in 12 frames over three different parts of the reciprocal space (36 frames total). Cell refinement and data reduction were performed with SAINT. Ver. 8.34A. (Bruker-AXS, 2014). Absorption correction was done by multiscan methods using SADABS Ver. 2015/1 ((Krause et al., 2015)). The structure was solved by direct methods and refined using SHELXL-2014 (Sheldrick, 2014).

To check the sample purity, a pattern matching of the X-ray powder diffraction experiment was performed using FullProf software. All the peaks were shown to match with the lattice parameters determined by single crystal X-ray diffraction.

### 4.3. Non-linear optical measurements

Non-Linear Optical (NLO) activity was evaluated by detecting the Second-Harmonic Generation (SHG) process. SHG measurements have been performed with an inverted microscope (Olympus IX71). The SHG process is induced by focusing in the sample a pulsed laser beam at 900 nm (Spectra Physics, Tsunami) with ultrashort pulse durations (100 fs at 80 MHz) and input power of 50 mW. Samples in powder form deposited onto a microscope slide are excited by this laser beam using a low-aperture microscope objective (Olympus, SLMPlan, X20, N.A. = 0.35). The SHG signal emitted at 450 nm is separated from the excitation beam by a dichroic mirror and collected during 1 s in reflection mode using a spectrometer (Acton research SP2300) coupled with a CCD camera (Princeton Instruments PIXIS400). For SHG imaging, a micrometric displacement stage is combined with a photon photomultiplier in counting mode (Hamamatsu, H7421).

**4.4. Magnetic and dielectric measurements**

The temperature ($T$) and magnetic field ($H$) dependent dc and ac magnetization ($M$) have been performed down to 2 K and up to 50 kOe using a commercial Superconducting Quantum Interference Device (SQUID) procured from Quantum Design (QD). Dielectric measurements were carried out using am impedance analyzer (LCR meter, E4886A, Agilent Technologies) with a home-made cryogenic insert which is integrated to Physical Properties Measurements System (PPMS) to sweep the magnetic field and temperature. We have analyzed the dielectric properties as a function of $T$ and $H$ for various frequencies ($\nu$) applying an ac voltage of 0.5 V amplitude in a powder sample in pellet form making a parallel plate capacitor with silver paint. The dielectric features have been reproduced by performing repeated experiments for different capacitors from different batch of samples and also tested with different ac voltage (say, 0.1-1 V) to confirm the intrinsic feature of the polycrystalline sample. All the temperature dependent physical measurements were performed during heating with a fixed temperature rate of 1 K/min in zero-field-cooled (ZFC) conditions, unless specifically stated in the manuscript. The features are also reproducible with different heating rate.

**Supporting Information**

**Designing of a magnetodielectric system in hybrid organic-inorganic framework, a perovskite layered phosphonate MnO$_3$PC$_6$H$_4$-*m*-Br.H$_2$O**

*Tathamay Basu\*, Clarisse Bloyet, Felicien Beaubras, Vincent Caignaert, Olivier Perez, Jean-Michel Rueff, Alain Pautrat and Bernard Raveau\**
Normandie Univ, ENSICAEN, UNICAEN, CNRS, CRISMAT, 6 Bd du Maréchal Juin, 14050 Caen Cedex, France.

\*E-mail: tathamaybasu@gmail.com; bernard.raveau@ensicaen.fr

*Jean-François Lohier*
Normandie Univ, ENSICAEN, UNICAEN, CNRS, LCMT, 6 Bd du Maréchal Juin, 14050 Caen Cedex, France.

*Paul-Alain Jaffrès, Helene Couthon*
CEMCA UMR CNRS 6521, Université de Brest, IBSAM, 6 Avenue Victor Le Gorgeu, 29238 Brest, France.

*Guillaume Rogez, Grégory Taupier and Honorat Dorkenoo*
IPCMS, UMR Unistra-CNRS 7504, 23 rue du Loess, BP 43, 67034, Strasbourg Cedex 2, France.

1. **Chemical characterisation of the precursor and ligand**

**Diethyl 3-bromophenylphosphonate :** $^{1}$**H NMR** (CDCl$_3$, 400 MHz): 1.33 (6H, t, $^{3}J$= 7.2 Hz, C$H_3$); 4.12 (4H, m, C$H_2$); 7.34 (1H, ddd, $^{3}J_{HH}$= 5.0 Hz, $^{3}J_{HH}$= 5.0 Hz, $^{4}J_{HP}$ = 12.4 Hz, C$H_{ar}$); 7.67 (1H, m, C$H_{ar}$); 7.73 (1H, dddd, $^{4}J_{HH}$= 1.0 Hz, $^{4}J_{HH}$= 1.0 Hz, $^{3}J_{HH}$= 7.6 Hz, $^{3}J_{HP}$ = 12.8 Hz, C$H_{ar}$); 7.94 (1H, ddd, $^{4}J_{HH}$= 1.0 Hz, $^{4}J_{HH}$= 2.0 Hz, $^{3}J_{HP}$ = 13.6 Hz, C$H_{ar}$) ; $^{31}$**P {$^{1}$H} NMR** (CDCl$_3$, 162 MHz): 17.1 ; $^{13}$**C{$^{1}$H} NMR** (CDCl$_3$, 125.75 MHz): 16.3 (C$H_3$-CH$_2$); 62.4 (C$H_2$-CH$_3$); 122.8 (d, $^{3}J_{PC}$= 19.8 Hz, $C$-Br); 130.1 (d, $^{2}J_{PC}$ = 12.6 Hz, C$H_{ar}$); 130.2 (d, $^{3}J_{PC}$= 5.9 Hz, C$H_{ar5}$; 131.1 (d, $^{1}J_{PC}$= 191.1 Hz, $C_{q1}$); 134.5 (d, $^{2}J_{PC}$ = 10.7 Hz, C$H_{ar2}$); 135.4 (d, $^{4}J_{PC}$ = 2.7 Hz, C$H_{ar}$) ; **m/z (ESI)**: 292.90 [M+H]$^{+}$; 586.95 [2M+H]$^{+}$ ; **IR** (cm$^{-1}$): 959 (C-O); 1015 (P-O); 1249 (P=O); 2982 (C-

**3-bromophenylphosphonic acid:** $^{1}$**H NMR** (CDCl$_3$, 400 MHz): 1.32 (12H, t, J = 6.8 Hz,O-C-CH$_3$); 2.60 (3H, s, Ar-CH$_3$); 4.12 (8H, m, -O-CH$_2$-); 7.68 (2H, m, H$_{Ar}$); 8.97 (1H, m, H$_{Ar}$); $^{31}$**P {$^{1}$H} NMR** (CDCl$_3$, 162 MHz): 14.73 ; $^{13}$**C{$^{1}$H} NMR** (CDCl$_3$, 125.75 MHz): δ 124.5 (d, $^{3}$J$_{PC}$= 17.1 Hz, Hz C-Br); 131.7 (d, J$_{PC}$= 8.3 Hz, C$_{Har}$); 132.7 (d, J$_{PC}$= 13.6 Hz, C$_{Har}$); 134.8 (s, C$_{Har}$); 135.6 (d, J$_{PC}$= 9.6 Hz, Hz C$_{Har}$); 144.8 (d, $^{1}$J$_{PC}$= 166 Hz, C-P).

**Table S1: atomic coordinates of Mn(H$_2$O)PO$_3$-C$_6$H$_4$-*m*-Br obtained at 150 K from single crystal X-ray refinements in the space group Pc**

|     | x / a      | y / b       | z / c        | Occ. | U$_{iso}$   |
| --- | ---------- | ----------- | ------------ | ---- | ----------- |
| Mn1 | 0.5988(8)  | 1.3429(11)  | 0.49891(16)  | 1    | 0.0131(11)  |
| Mn2 | 0.0995(8) | 0.8396(11) | 0.51385(15) | 1 | 0.0129(11) |
| P2 | 0.1585(15) | 0.3373(18) | 0.5615(3) | 1 | 0.0118(16) |
| P1 | 0.6613(15) | 0.8502(18) | 0.4512(3) | 1 | 0.0109(16) |
| O2 | 0.802(4) | 0.608(4) | 0.4716(7) | 1 | 0.010(4) |
| O4 | 0.307(4) | 0.135(5) | 0.5415(7) | 1 | 0.017(5) |
| O3 | -0.134(4) | 0.330(5) | 0.5541(7) | 1 | 0.016(5) |
| O0AA | 0.804(5) | 1.037(5) | 0.4691(7) | 1 | 0.019(5) |
| O2AA | 0.294(4) | 0.563(5) | 0.5415(7) | 1 | 0.013(5) |
| O1AA | 0.368(5) | 0.858(5) | 0.4593(7) | 1 | 0.020(5) |
| C2 | 0.935(5) | 0.959(4) | 0.3285(7) | 1 | 0.023(8) |
| C3 | 0.880(4) | 0.985(4) | 0.3724(7) | 1 | 0.016(7) |
| H3 | 0.94719 | 1.11548 | 0.38750 | 1 | 0.0190 |
| C4 | 0.727(5) | 0.818(5) | 0.3940(6) | 1 | 0.013(6) |
| C5* | 0.629(6) | 0.625(5) | 0.3717(9) | 1 | 0.059(15) |
| H5* | 0.52426 | 0.51164 | 0.38644 | 1 | 0.0710 |
| C6* | 0.684(6) | 0.600(5) | 0.3279(9) | 1 | 0.064(17) |
| H6* | 0.61688 | 0.46906 | 0.31277 | 1 | 0.0770 |
| C7 | 0.837(5) | 0.767(5) | 0.3063(6) | 1 | 0.024(8) |
| H7 | 0.87488 | 0.74953 | 0.27645 | 1 | 0.0290 |
| C1* | 0.123(6) | 0.137(5) | 0.6413(8) | 1 | 0.052(13) |
| H1* | 0.01289 | 0.02696 | 0.62674 | 1 | 0.0630 |
| C8* | 0.179(6) | 0.110(5) | 0.6850(8) | 1 | 0.069(18) |
| H8_ | 0.10749 | -0.01834 | 0.70024 | 1 | 0.0830 |
| C9 | 0.341(5) | 0.272(5) | 0.7064(6) | 1 | 0.017(7) |
| H9 | 0.37973 | 0.25387 | 0.73620 | 1 | 0.0200 |
| C10 | 0.447(5) | 0.461(5) | 0.6840(7) | 1 | 0.027(9) |
| C11 | 0.391(5) | 0.488(4) | 0.6403(7) | 1 | 0.018(7) |
| H11 | 0.46289 | 0.61668 | 0.62506 | 1 | 0.0220 |
| C12 | 0.229(5) | 0.326(5) | 0.6189(5) | 1 | 0.014(7) |
| Br1* | 1.1932(12) | 1.1543(15) | 0.30132(16) | 1 | 0.072(3) |
| Br2* | 0.6952(12) | 0.6468(15) | 0.71157(16) | 1 | 0.073(3) |
| O1 | -0.230(4) | 0.838(5) | 0.5626(7) | 1 | 0.015(5) |
| H1A | -0.22955 | 0.91612 | 0.58989 | 1 | 0.01(9) |
| H1B | -0.41342 | 0.78219 | 0.55955 | 1 | 0.00(8) |
| O5 | 0.270(5) | 1.353(5) | 0.4489(8) | 1 | 0.019(5) |
| H5A | 0.23399 | 1.19286 | 0.45406 | 1 | 0.00(8) |
| H5B | 0.09226 | 1.42079 | 0.44529 | 1 | 0.00(8) |

Note : the atoms C* and H* (C1, H1, C5,H5, C6, H6,C8, H8) and Br* exhibit abnormally high thermal factors due to the fact that the "C$_6$H$_4$Br" rings in the organic layers, show a high disordering in their various orientations with respect to each other. Consequently, the

corresponding coordinates of these atoms cannot be considered as significant and are responsible for the rather high agreement factors obtained from these refinements.

**Table S2 : Anisotropic atomic displacement parameters (Å²)**

|     | $U_{11}$ | $U_{22}$ | $U_{33}$ | $U_{12}$ | $U_{13}$ | $U_{23}$ |
| --- | --- | --- | --- | --- | --- | --- |
| Br1 | 0.084(4) | 0.101(6) | 0.031(3) | -0.066(4) | 0.011(2) | 0.005(3) |
| Br2 | 0.081(4) | 0.104(7) | 0.034(3) | -0.071(4) | -0.012(2) | 0.004(3) |

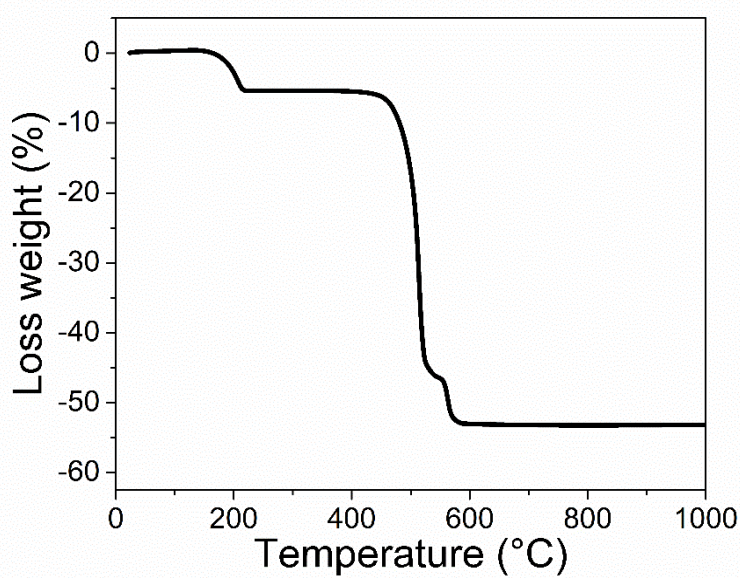

**Figure S1**: Thermogravimetric Analysis curves of Mn(H$_2$O)PO$_3$-C$_6$H$_4$-*m*-Br compound recorded under air at a heating rate of 3°C/min.

**Magnetic and dielectric properties of Mn(H$_2$O)PO$_3$-C$_6$H$_4$:**

The magnetic results are depicted in Figure S2 for the compound Mn(H$_2$O)PO$_3$-C$_6$H$_5$. The magnetic behavior is same as Mn(H$_2$O)PO$_3$-C$_6$H$_4$-*m*-Br compound, that is, a broad peak ~20 K ($T_0$) followed by complex magnetic ordering $T_1$~11.6 K. The slight change in $T_1$ is probably due to small change in inter-planer distance, as discussed in main manuscript (also see Ref. 39-40). The Curie-Weiss fit of the inverse susceptibility for $T > 150$ K yields an effective moment of $\mu_{eff} = 5.99\ \mu_B$ and a Curie-Weiss temperature of $\Theta_{CW} = -37.9$ K, with a temperature independent contribution of $\chi_0 = -1.1*10^{-4}$ emu/mol. The evolution of $T_1$ with different magnetic field is nearly same for both the compound.

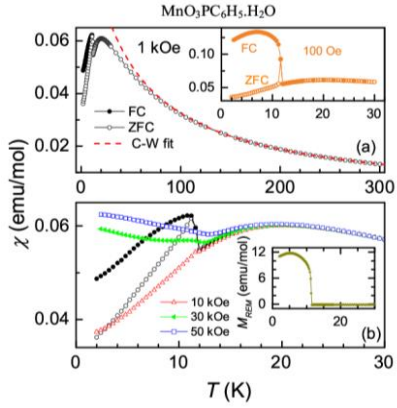

**Figure S2:** (a) Dc susceptibility as a function of temperature for 1 kOe magnetic field for zero-field-cooled and field-cooled condition from 2-300 K for the compound $MnO_3PC_6H_5 \cdot H_2O$; The Currie-Weiss fitting in paramagnetic region (150-300 K). Inset of (a): Dc susceptibility as a function of T for 100 Oe for ZFC and FC conditions from 2-30 K. (b) T-dependent dc susceptibility in presence of different magnetic field from 2-30 K. Inset of (b): Remnant magnetization is plotted as a function of *T*.

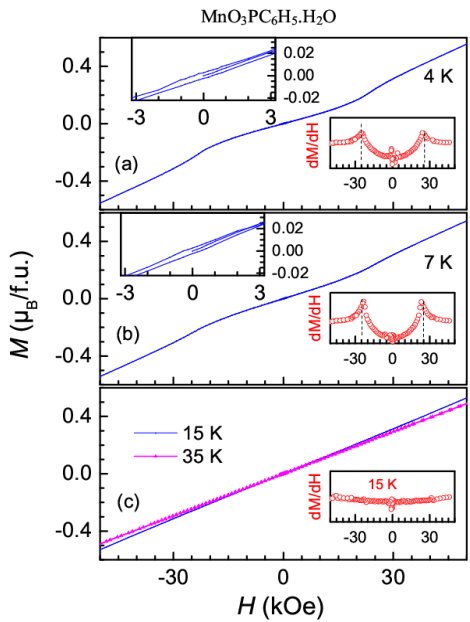

**Figure S3:** Isothermal magnetization as function of magnetic field for *T*= 2 K (a), 7 K (b), 15 and 35 K (c), for the compound $MnPO_3C_6H_5 \cdot H_2O$. Right inset of (a), (b) and (c) shows *dM/dH* as a function of *H* for 2, 7 and 15 K, respectively, to highlight the metamagnetic transitions around 25 kOe for *T<T_1*. Left inset of (a) and (b) shows the enlarged version of *M(H)* at low magnetic field to highlight the weak hysteresis at 4 and 7 K, respectively.

The isothermal magnetization at different temperature is plotted in Figure S3 for selective temperature. A clear *H*-induced transition is observed ~27 kOe for this compound for $T<T_1$. The small hysteresis loop (see left inset of Figure S3), continuous linear increase of M(H) and low value of magnetization at 50 kOe high magnetic field confirms weak ferromagnetic (canted AFM) behavior for this compound as well, as observed in the titled compound of this manuscript. Therefore, the magnetic features are similar for both the compounds (Mn(H$_2$O)PO$_3$-C$_6$H$_5$ and Mn(H$_2$O)PO$_3$-C$_6$H$_4$-*m*-Br) in this system, which suggest that Br-does not have any significant role on magnetism, which is quite usual.

The real part of dielectric constant $\varepsilon'$ is depicted as a function of temperature in Figure S4 for the compound Mn(H$_2$O)PO$_3$-C$_6$H$_5$. We did not observe any clear feature at the onset of magnetic ordering ($T_0$ and $T_1$) apart from small change in slope, within the resolution of the experiment, unlike Mn(H$_2$O)PO$_3$-C$_6$H$_4$-*m*-Br. As discussed in the main manuscript, this is attributed to weak nature of dipolar strength or/and absence of dipolar ordering, and thereby weak (negligible) ME coupling, despite of the fact that this system itself could be prone for multiferroic and ME coupling.

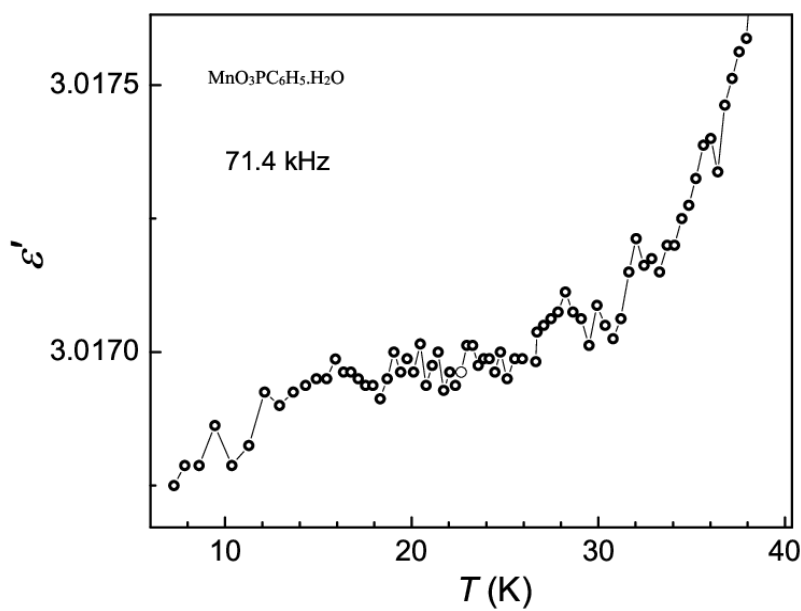

**Figure S4:** Real part of dielectric constant as a function of temperature for a fixed frequency of 71.4 kHz for the compound MnO$_3$PC$_6$H$_5$.H$_2$O.

**Acknowledgements**

The authors thank the Agence Nationale de la Recherche [contract no. ANR-14-CE07-0004-01 (HYMN)] for financial support. The authors also express their grateful acknowledgment for technical support to Sylvie Collin and Fabien Veillon from the CRISMAT laboratory, CNRS et ENSICAEN, Caen, France and to Sylvie Hernot form the CEMCA laboratory and the Service de RMN, UFR Sciences et Techniques, Université de Brest, France.

**References:**

[1] Y. Tokura, S. Seki, N. Nagaosa, *Rep. Prog. Phys.* **2014**, *77*, 076501.
[2] N. A. Spaldin, M. Fiebig, *Science* **2005**, *309*, 391.
[3] D. Khomskii, *Physics* **2009**, *2*.
[4] W. Eerenstein, N. D. Mathur, J. F. Scott, *Nature* **2006**, *442*, 759.
[5] S. Horiuchi, Y. Tokunaga, G. Giovannetti, S. Picozzi, H. Itoh, R. Shimano, R. Kumai, Y. Tokura, *Nature* **2010**, *463*, 789.
[6] M. Bibes, A. Barthélémy, *Nat. Mater.* **2008**, *7*, 425.
[7] J. T. Heron, D. G. Schlom, R. Ramesh, *Appl. Phys. Rev.* **2014**, *1*, 021303.
[8] R. Ramesh, *Nature* **2009**, *461*, 1218.
[9] G. Rogez, N. Viart, M. Drillon, *Angew. Chem. Int. Ed.* **n.d.**, *49*, 1921.
[10] H. Cui, Z. Wang, K. Takahashi, Y. Okano, H. Kobayashi, A. Kobayashi, *J. Am. Chem. Soc.* **2006**, *128*, 15074.
[11] H.-B. Cui, K. Takahashi, Y. Okano, H. Kobayashi, Z. Wang, A. Kobayashi, *Angew. Chem. Int. Ed.* **n.d.**, *44*, 6508.
[12] P. Jain, V. Ramachandran, R. J. Clark, H. D. Zhou, B. H. Toby, N. S. Dalal, H. W. Kroto, A. K. Cheetham, *J. Am. Chem. Soc.* **2009**, *131*, 13625.
[13] P. Jain, N. S. Dalal, B. H. Toby, H. W. Kroto, A. K. Cheetham, *J. Am. Chem. Soc.* **2008**, *130*, 10450.
[14] Y. Tian, A. Stroppa, Y. Chai, L. Yan, S. Wang, P. Barone, S. Picozzi, Y. Sun, *Sci. Rep.* **2014**, *4*, 6062.
[15] Y. Tian, S. Shen, J. Cong, L. Yan, S. Wang, Y. Sun, *J. Am. Chem. Soc.* **2016**, *138*, 782.
[16] W. Wang, L.-Q. Yan, J.-Z. Cong, Y.-L. Zhao, F. Wang, S.-P. Shen, T. Zou, D. Zhang, S.-G. Wang, X.-F. Han, Y. Sun, *Sci. Rep.* **2013**, *3*, 2024.
[17] A. Stroppa, P. Jain, P. Barone, M. Marsman, J. M. Perez-Mato, A. K. Cheetham, H. W. Kroto, S. Picozzi, *Angew. Chem. Int. Ed.* **n.d.**, *50*, 5847.
[18] A. Stroppa, P. Barone, P. Jain, J. M. Perez-Mato, S. Picozzi, *Adv. Mater.* **n.d.**, *25*, 2284.
[19] L. C. Gómez-Aguirre, B. Pato-Doldán, J. Mira, S. Castro-García, M. A. Señarís-Rodríguez, M. Sánchez-Andújar, J. Singleton, V. S. Zapf, *J. Am. Chem. Soc.* **2016**, *138*, 1122.
[20] Z.-J. Lin, J. Lü, M. Hong, R. Cao, *Chem. Soc. Rev.* **2014**, *43*, 5867.
[21] T. Basu, A. Jesche, B. Bredenkötter, M. Grzywa, D. Denysenko, D. Volkmer, A. Loidl, S. Krohns, *Mater. Horiz.* **2017**, *4*, 1178.

[22] E. Pardo, C. Train, H. Liu, L.-M. Chamoreau, B. Dkhil, K. Boubekeur, F. Lloret, K. Nakatani, H. Tokoro, S. Ohkoshi, M. Verdaguer, *Angew. Chem. Int. Ed.* **n.d.**, *51*, 8356.
[23] A. O. Polyakov, A. H. Arkenbout, J. Baas, G. R. Blake, A. Meetsma, A. Caretta, P. H. M. van Loosdrecht, T. T. M. Palstra, *Chem. Mater.* **2012**, *24*, 133.
[24] B. Kundys, A. Lappas, M. Viret, V. Kapustianyk, V. Rudyk, S. Semak, Ch. Simon, I. Bakaimi, *Phys. Rev. B* **2010**, *81*, 224434.
[25] Q. Evrard, Z. Chaker, M. Roger, C. M. Sevrain, E. Delahaye, M. Gallart, P. Gilliot, C. Leuvrey, J.-M. Rueff, P. Rabu, C. Massobrio, M. Boero, A. Pautrat, P.-A. Jaffrès, G. Ori, G. Rogez, *Adv. Funct. Mater.* **n.d.**, *27*, 1703576.
[26] G. Cao, H. Lee, V. M. Lynch, T. E. Mallouk, *Inorg. Chem.* **1988**, *27*, 2781.
[27] J. Le Bideau, C. Payen, B. Bujoli, P. Palvadeau, J. Rouxel, *J. Magn. Magn. Mater.* **1995**, *140–144*, 1719.
[28] S. G. Carling, P. Day, D. Visser, R. K. Kremer, *J. Solid State Chem.* **1993**, *106*, 111.
[29] C. M. Sevrain, M. Berchel, H. Couthon, P.-A. Jaffrès, *Beilstein J. Org. Chem.* **2017**, *13*, 2186.
[30] J.-M. Rueff, V. Caignaert, S. Chausson, A. Leclaire, C. Simon, O. Perez, L. L. Pluart, P.-A. Jaffrès, *Eur. J. Inorg. Chem.* **2008**, *2008*, 4117.
[31] P. Tavs, *Chem. Ber.* **1970**, *103*, 2428.
[32] J.-M. Rueff, N. Barrier, S. Boudin, V. Dorcet, V. Caignaert, P. Boullay, G. B. Hix, P.-A. Jaffrès, *Dalton Trans.* **2009**, *0*, 10614.
[33] N. Hugot, M. Roger, J.-M. Rueff, J. Cardin, O. Perez, V. Caignaert, B. Raveau, G. Rogez, P.-A. Jaffres, *Eur. J. Inorg. Chem.* **2015**, *2016*, 266.
[34] O. Perez, C. Bloyet, J.-M. Rueff, N. Barrier, V. Caignaert, P.-A. Jaffrès, B. Raveau, *Cryst. Growth Des.* **2016**, *16*, 6781.
[35] J.-M. Rueff, V. Caignaert, A. Leclaire, C. Simon, J.-P. Haelters, P.-A. Jaffrès, *CrystEngComm* **2009**, *11*, 556.
[36] J.-M. Rueff, O. Perez, A. Pautrat, N. Barrier, G. B. Hix, S. Hernot, H. Couthon-Gourvès, P.-A. Jaffrès, *Inorg. Chem.* **2012**, *51*, 10251.
[37] G. A. Bain, J. F. Berry, *J. Chem. Educ.* **2008**, *85*, 532.
[38] V. Laget, C. Hornick, P. Rabu, M. Drillon, R. Ziessel, *Coord. Chem. Rev.* **1998**, *178–180*, 1533.
[39] M. Drillon, P. Panissod, *J. Magn. Magn. Mater.* **1998**, *188*, 93.
[40] P. Rabu, S. Rouba, V. Laget, C. Hornick, M. Drillon, *Chem. Commun.* **1996**, *0*, 1107.
[41] Y. Tian, W. Wang, Y. Chai, J. Cong, S. Shen, L. Yan, S. Wang, X. Han, Y. Sun, *Phys. Rev. Lett.* **2014**, *112*, 017202.
[42] K. Singh, T. Basu, S. Chowki, N. Mahapotra, K. K. Iyer, P. L. Paulose, E. V. Sampathkumaran, *Phys. Rev. B* **2013**, *88*, 094438.
[43] T. Basu, C. Bloyet, J.-M. Rueff, V. Caignaert, A. Pautrat, B. Raveau, G. Rogez, P.-A. Jaffrès, *J. Mater. Chem. C* **2018**, *6*, 10207.
[44] T. Basu, V. V. R. Kishore, S. Gohil, K. Singh, N. Mohapatra, S. Bhattacharjee, B. Gonde, N. P. Lalla, P. Mahadevan, S. Ghosh, E. V. Sampathkumaran, *Sci. Rep.* **2014**, *4*, 5636.
[45] G. Nénert, T. T. M. Palstra, *Phys. Rev. B* **2007**, *76*, 024415.
[46] T. Basu, K. K. Iyer, K. Singh, E. V. Sampathkumaran, *Sci. Rep.* **2013**, *3*, 3104.
[47] T. Basu, D. T. Adroja, F. Kolb, H.-A. Krug von Nidda, A. Ruff, M. Hemmida, A. D. Hillier, M. Telling, E. V. Sampathkumaran, A. Loidl, S. Krohns, *Phys. Rev. B* **2017**, *96*, 184431.
[48] X. Wan, H.-C. Ding, S. Y. Savrasov, C.-G. Duan, *Sci. Rep.* **2016**, *6*, 22743.
[49] N. Hur, S. Park, P. A. Sharma, J. S. Ahn, S. Guha, S.-W. Cheong, *Nature* **2004**, *429*, 392.

[50] G. R. Blake, L. C. Chapon, P. G. Radaelli, S. Park, N. Hur, S.-W. Cheong, J. Rodríguez-Carvajal, *Phys. Rev. B* **2005**, *71*, 214402.
[51] V. Balédent, S. Chattopadhyay, P. Fertey, M. B. Lepetit, M. Greenblatt, B. Wanklyn, F. O. Saouma, J. I. Jang, P. Foury-Leylekian, *Phys. Rev. Lett.* **2015**, *114*, 117601.
[52] Y. Zhang, W.-Q. Liao, D.-W. Fu, H.-Y. Ye, C.-M. Liu, Z.-N. Chen, R.-G. Xiong, *Adv. Mater.* **2015**, *27*, 3942.
[53] L. Brammer, E. A. Bruton, P. Sherwood, *Cryst. Growth Des.* **2001**, *1*, 277.
[54] C. L. D. Gibb, E. D. Stevens, B. C. Gibb, *J. Am. Chem. Soc.* **2001**, *123*, 5849.
[55] D. Wang, J. Wang, D. Zhang, Z. Li, *Sci. China Chem.* **2012**, *55*, 2018.